**Making Waves: Defining the lead time of wastewater-based epidemiology for COVID-19**

Scott W. Olesen, Maxim Imakaev, Claire Duvallet*

Biobot Analytics, Inc., Cambridge, MA, USA

* Correspondence: claire@biobot.io

**Abstract** (161 words)

Individuals infected with SARS-CoV-2, the virus that causes COVID-19, may shed the virus in stool before developing symptoms, suggesting that measurements of SARS-CoV-2 concentrations in wastewater could be a "leading indicator" of COVID-19 prevalence. Multiple studies have corroborated the leading indicator concept by showing that the correlation between wastewater measurements and COVID-19 case counts is maximized when case counts are lagged. However, the meaning of "leading indicator" will depend on the specific application of wastewater-based epidemiology, and the correlation analysis is not relevant for all applications. In fact, the quantification of a leading indicator will depend on epidemiological, biological, and health systems factors. Thus, there is no single "lead time" for wastewater-based COVID-19 monitoring. To illustrate this complexity, we enumerate three different applications of wastewater-based epidemiology for COVID-19: a qualitative "early warning" system; an independent, quantitative estimate of disease prevalence; and a quantitative alert of bursts of disease incidence. The leading indicator concept has different definitions and utility in each application.

**Wastewater-based epidemiology as a "leading indicator"**

Wastewater-based epidemiology (WBE), the use of measurements from wastewater for public health surveillance, is being used in the COVID-19 pandemic as a complement to more traditional monitoring methods like diagnostic testing (National Wastewater Surveillance System 2020; Larsen & Wigginton 2020; Polo et al. 2020). WBE gained particular attention in part because wastewater concentrations of SARS-CoV-2 have been characterized as a "leading indicator" of reported COVID-19 case counts (Table 1 and references therein; Keshaviah et al. 2021). The biological principle behind wastewater as a leading indicator is that many infected individuals shed the virus in stool before they develop symptoms and thus also before they seek medical care (Daughton 2020; Zhu et al. 2021).

However, we suspect the term "leading indicator" is being used in multiple senses in the context of COVID-19 WBE, perhaps because the term, originally used in economics and business (Bloom et al. 2007), has not seen widespread use in infectious disease public health.[1] To explore what it means for wastewater to be a "leading indicator" for COVID-19 and how wastewater's lead time can be quantified, we review three main applications of WBE for COVID-19 and discuss what "leading indicator" means in the context of each application.

**Application 1 - Qualitative detection of disease presence/absence**

The first application is a qualitative "early warning" system, testing for a detection or nondetection of SARS-CoV-2 in wastewater (Daughton 2020; Hassard et al. 2020; National Wastewater Surveillance System 2020; Zhu et al. 2021). In other words, the goal in this application is to answer the question: are there currently more than zero infected individuals in the surveilled population?

---

[1] A PubMed search on 10 Mar 2021 for "leading indicator" and any of "public health", "infectious disease", "SARS-CoV-2", or "COVID-19" yielded 38 publications, many of which dealt with occupational health, predictors of an individual's disease trajectory, or "leading" in the sense of "most important" rather than "temporally before" (Table 1).

This application is relevant in the context of individual facilities, such as a college dormitory, correctional institution, nursing home, or cruise ship (Ahmed et al. 2020b; Betancourt et al. 2020; Daughton 2020; Harris-Lovett et al. 2021; Peiser 2020; Reeves et al. 2021; Targeted Wastewater Surveillance at Facilities, Institutions, and Workplaces 2020), as well as in the context of larger wastewater catchments, such as a city with little ongoing transmission of SARS-CoV-2 (Ahmed et al. 2020a; COVID-19 weekly surveillance reports; Fongaro et al. 2021; Jørgensen et al. 2020; Medema et al. 2020; Randazzo et al. 2020). If the surveilled population has zero (or very few) known cases but does have detectable SARS-CoV-2 in wastewater, then there are likely one or more individuals in the population with presymptomatic or asymptomatic infections. WBE's lead time in this application is the delay between detection of the virus in wastewater versus the detection of cases by other means, such as if the infected person is identified with a screening test, or if an infected person becomes symptomatic, seeks a diagnostic test, and receives a positive result. Ideally, this lead time allows for the implementation of mitigation measures, like quarantine or mass diagnostic testing, that can prevent an outbreak.

In this application, WBE's lead time depends on biological and health systems factors. In terms of biology, WBE's lead time depends on the proportion of infected people who shed detectable levels of virus in their stool (Jones et al. 2020) as well as the time delay between the onset of viral shedding in feces versus the onset of symptoms (Miura et al. 2021). The lead time also depends on health systems factors like the availability of testing, individuals' healthcare seeking behavior, and the turnaround time for returning diagnostic test results and wastewater monitoring measurements (McClary-Gutierrez et al. 2021). For example, if there are sufficient resources to allow screening everyone in the monitored population for SARS-CoV-2 every day, then wastewater's lead time and added value will be minimal (Peccia et al. 2020). Conversely, if there is no active case finding in the relevant population, then wastewater, insofar as it can detect the presence of infected individuals before they present with symptoms, could be a "leading indicator" with a lead time at least as long as the delay between onset of fecal shedding and the onset of symptoms, that is, 1 to 6 days (Zhu et al. 2021). If the turnaround time for diagnostic testing is longer than the turnaround time for wastewater testing, wastewater's lead time will be that much greater.

In theory, the biological factors affecting WBE's lead time could include viral lineage. For example, infection with different variants could lead to different presymptomatic shedding patterns. To our knowledge, however, there are no studies that have investigated differences in shedding dynamics by virus variant.

**Application 2 - Independent, quantitative estimate of community-level disease prevalence and trends**

The second application of WBE for COVID-19 is as a quantitative, population-level estimate of disease prevalence. Rather than estimating SARS-CoV-2 prevalence using just those individuals who tested positive and were formally counted as a case, wastewater detects virus in a sample formed from the pooled excretions of many individuals (Jones et al. 2020; Kitajima et al. 2020), regardless of whether they are symptomatic, have access to healthcare, or whose healthcare system has abundant resources for testing (Medema et al. 2020). Thus, this application is appealing because it provides estimates that are potentially less biased and less resource-intensive compared to traditional disease monitoring using diagnostic testing (National Wastewater Surveillance System 2020; D'Aoust et al. 2021; Daughton 2020; Larsen & Wigginton 2020; Wu et al. 2020a).

WBE and diagnostic testing can also provide synergistic insights. If trends in both case rates and wastewater change directions, public health officials can be more certain that disease prevalence has truly passed an inflection point. If wastewater measurements rise while case counts remain stable, then diagnostic testing may be undercounting cases (Fernandez-Cassi et al. 2021; Wu et al. 2020a). Conversely, if rates of positive diagnostic tests rise but wastewater measurements remain stable, then the apparent increase in cases may be due to increased testing (i.e., less undercounting) rather than to increased prevalence (Gerrity et al. 2021; McClary-Gutierrez et al. 2021).

In principle, this second application of WBE could be a "leading indicator". Various reports have suggested that correlations between wastewater concentrations of SARS-CoV-2 and reported COVID-19 cases are maximized when cases are lagged by 2 to 8 days (i.e., when wastewater leads cases by 2 to 8 days; Table 1). This interpretation, however, is subject to three important caveats.

First, to say that wastewater is a "leading indicator" with a lead time of $D$ days does not mean that today's wastewater measurement says exactly how many new cases will be reported $D$ days from now. Of course, neither wastewater measurements nor reported case counts are perfectly accurate: the time series of new case counts, offset by $D$ days, would never perfectly line up with the timeseries of WBE measurements because of measurement error (Gerrity et al. 2021). However, even if wastewater measurements and reported case counts were perfectly accurate, we would not expect the two time series to line up, since individuals who are shedding virus in stool today are at different places in their disease trajectories (Hoffman & Alsing 2021; Miura et al. 2021). Some will become symptomatic sooner, some later, and some not at all. Some may already have become counted as reported cases. In other words, there are multiple factors that delay cases counts relative to the onset of fecal shedding: disease time courses, healthcare seeking behavior, access to testing, and testing turnaround time. These delays mean that the timeseries of new case counts is offset relative to WBE measurements. However, there is also variability in those factors, which leads to a smoothing of case counts relative to shedding rates (Fernandez-Cassi et al. 2021; Wu et al. 2020b). This combination of shifting and smoothing means that the temporal relationship between wastewater measurements and case counts is complex (Weidhaas et al. 2021).

Second, even insofar as a single lead time is meaningful for two signals that do not perfectly align, that lead time will vary across populations and through time in a single population, and it will vary depending on the statistical methodology used (Table 1, Figure 1). From a biological point of view, we would expect the lead time to change depending on who is being infected, their disease severity, and the precise SARS-CoV-2 variant that infected them (Cevik et al. 2020; Kissler et al. 2021). From a health systems point of view, the delay between shedding virus and becoming a reported case depends on the availability and accessibility of diagnostic tests. More testing and faster turnaround time means shorter lead time. Thus, we should not expect that studies will all report the same lead time for this application of WBE for COVID-19.

Third, even accounting for all the variability mentioned above, there may be many lead times that describe the relationship between WBE measurements and reported case counts nearly equally well (Figure 1; D'Aoust et al. 2021; Graham et al. 2021). Thus, it is not the case that wastewater signals are well correlated only with case rates exactly $D$ days in the future; instead, wastewater correlates positively with case rates over many days in the future. Asserting that the "best" lead time is the one that maximizes this correlation thus ignores other factors important to public health practitioners. It might be preferable, for example, to maximize a practitioner's early warning by having a slightly poorer prediction of case rates at a greater lead time than the numerically optimal prediction at a shorter lead time.

Given the complexity around wastewater's lead time in this application, we hypothesize that WBE has more value as an independent measurement of population-level disease prevalence rather than as a precise leading indicator of prevalence (Reeves et al. 2021). As an analogy, we note that influenza monitoring systems include counts of laboratory-confirmed influenza-related hospitalizations as well as counts of outpatient visits for influenza-like illness. In principle, outpatient visits could be a leading indicator of hospitalizations because individuals with severe disease would visit a physician before becoming hospitalized. However, the main value of outpatient visits for influenza monitoring is not as a leading indicator but rather as a complementary window onto disease prevalence. We speculate that a few days' extra notice is not as important as having an independent confirmation that flu season is generally accelerating, or has peaked and is on the decline. Analogously, for this application, WBE's primary value is in providing an independent confirmation that COVID-19 prevalence is generally high or low, or generally rising or falling, rather than having a lead time of 2 to 8 days (Gonzalez et al. 2020).

In resource-limited settings with very limited testing, where wastewater-based testing could in theory be the primary mechanism for disease prevalence monitoring (Hart & Halden 2020), this logic would hold even more strongly: WBE's primary value would be in measuring COVID-19 prevalence, rather than in being a leading indicator relative to the limited number of diagnostic tests.

**Application 3 - Quantitative estimate of rapid changes in disease incidence**
The third application is to detect a rapid change, or "burst", in disease incidence using wastewater measurements on a background of ongoing transmission. For example, given daily wastewater sampling through a holiday or special event, such as Thanksgiving or the

Superbowl in the US, how many excess infections occurred specifically because of that event? This information could help public health officials estimate the timing and magnitude of the resulting surge in cases and hospitalizations following those events. As another example, given regular wastewater sampling at a congregate living facility where convalescent shedders from a previous outbreak have been identified and isolated, can a second, unrelated outbreak be rapidly detected and distinguished from the first outbreak?

In this application, WBE would be a "leading indicator" insofar as it would provide an alert about the burst in disease activity before other monitoring systems, like case counts, would. However, to our knowledge, this application has seen only anecdotal and not quantitatively rigorous use (Fox 2020), likely because it is substantially more complex than the first two. Unlike Application 1, this application must deliver a quantitative estimate of disease incidence (i.e., the change in the number of infected individuals), not just a simple presence/absence determination. Unlike Application 2, which involves identifying trends in disease activity over weeks, this application requires quantifying *changes in trends* that occur rapidly, perhaps on the timescale of days rather than weeks. This short timescale exacerbates the analytical challenges faced by the other applications, including the variability in wastewater measurements and case counts.

Validating this application will be especially challenging, as it would require establishing a gold standard of true disease incidence to compare its predictions against. Case counts, although an attractive indicator of disease activity, cannot themselves be the gold standard because of their weekly patterns (e.g., fewer cases counted on weekends) and their unusual behavior during holidays and special events (e.g., the anomalous case count patterns during the major US holidays in November and December 2020). Rolling 7-day averages of case counts are also an inappropriate gold standard. Although weekly rolling averages are not subject to within-week variations in case reporting, they obscure exactly the short-term bursts in incidence this application aims to detect (Bloom et al. 2007). Thus, a robust burst-detection algorithm likely needs to include a sophisticated statistical inference of underlying disease incidence, which remains an area of ongoing research (Li et al. 2021).

**Conclusions**

- There are at least three distinct applications of WBE for COVID-19: (1) qualitatively detecting new infections on a background of little or no disease activity, (2) independently estimating COVID-19 prevalence and trends, and (3) detecting sudden increases in disease incidence.
- WBE can be a "leading indicator", but there is no one single "lead time" because the lead time depends on the application of WBE, and it depends on biological, epidemiological, and health systems factors.
- For Application 1, wastewater is a leading indicator insofar as it can identify new infections before those individuals would be identified by some other method. If there are sufficient resources to allow screening everyone in the monitored population for SARS-CoV-2 every day, then wastewater likely leads reported case counts by very little time, if any. If there is no active case finding, then WBE is a leading indicator of at least 2 to 8 days, that is, the time between when an individual begins shedding detectable virus in their stool and when they seek diagnostic testing.
- The primary value of Application 2 may be that WBE measurements are unaffected by access to testing, healthcare-seeking behaviors, or other socio-behavioral factors. Thus, for this application, WBE may be more valuable because it is an independent indicator of disease prevalence, rather than because it can be a leading indicator.
- Application 3, detecting sudden increases in disease incidence on a background of high disease prevalence, will pose substantial technical challenges.

- Practitioners should recognize that the utility of WBE for public health will vary between applications and will change over time as our methodologies improve and as the epidemiology of COVID-19 changes. Lead times are only one part of that larger picture.

**Acknowledgments**

We thank Eric Alm, Mary Bushman, Nour Sharara, and Amy Xiao for helpful discussions.

**Statements**

This research did not receive any specific grant from funding agencies in the public, commercial, or not-for-profit sectors. SO, MI, and CD are employees of Biobot Analytics, Inc.

**Table 1.** Studies that made a quantitative estimate of the lead times for wastewater SARS-CoV-2 concentrations in the context of Application 2 (i.e., community-wide prevalence estimation). Studies that report an "early warning" approach but do not make a quantitative estimate of WBE's lead time are not included.

| Study | Study location | Study period | Lead time (days) | Methodology for quantifying lead time |
|---|---|---|---|---|
| D'Aoust et al. 2021 | Ottawa, Canada | Jun-Aug 2020 | 2 | Maximum Pearson correlation between wastewater and number of new cases |
| Feng et al. 2021 | 12 WWTPs covering 10 cities in Wisconsin, USA | Aug 2020-Jan 2021 | 0 to 6 (different for each WWTP) | Maximum Spearman correlation between wastewater and smoothed (7-day average) number of new cases |
| Kumar et al. 2021 | Gandhinagar, Gujarat, India | Aug-Sep 2020 | 14 | Visual inspection of percent change in wastewater concentration and number of new cases |
| Larsen et al. 2021 | 24 WWTPs in upstate New York, USA | May-Dec 2020 | 3 (active cases); 6 (incidence) | Maximum Pearson correlation between wastewater measurements and "active cases" (sum of cases over past 10 days) or incidence (7-day average of new cases) |
| Nemudryi et al. 2020 | Bozeman, Montana, USA | Mar-Jun 2020 | 2 (mid-March); 4 (May) | Maximum Pearson correlation between wastewater measurements and number of positive tests |
| Peccia et al. 2020 | New Haven, Connecticut, USA | Mar-Jun 2020 | 6-8 | Distributed lag time series model linking wastewater measurements and number of positive tests by report date |
| Wu et al. 2020b | Greater Boston, MA, USA | Jan-May 2020 | 4 (maximum correlation); 4-10 (range) | Pearson correlation between unsmoothed viral titers in wastewater and number of new cases |
| Wurtzer et al. 2020 | Paris, France | Mar 2020 | 8 | Visual inspection of wastewater and number of positive tests |

**Figure 1.** Correlations between measured wastewater SARS-CoV-2 concentrations and reported COVID-19 case counts (y-axis) vary depending on lead time (x-axis), correlation metric (panel a), incidence metric (b), and time (c) in Boston, Massachusetts during April 2020 to March 2021. Wastewater data were collected using methods previously described (Wu et al. 2020a; Wu et al. 2020b; data available at https://www.mwra.com/biobot/biobotdata.htm, North system). Case data includes cases in Suffolk and Middlesex Counties, MA, which are served by the wastewater plant (USA Facts; data available at https://usafacts.org/visualizations/coronavirus-covid-19-spread-map). Zero lead time refers to the correlation between wastewater and the case counts on the day of the wastewater sampling. Positive lead times refer to wastewater correlated with later case counts (e.g., a lead time of +3 days refers to the correlation between wastewater and cases 3 days later). Negative lead times refer to wastewater correlated with earlier case counts. Colors are only used to distinguish curves; the same color in different subplots are not necessarily related.

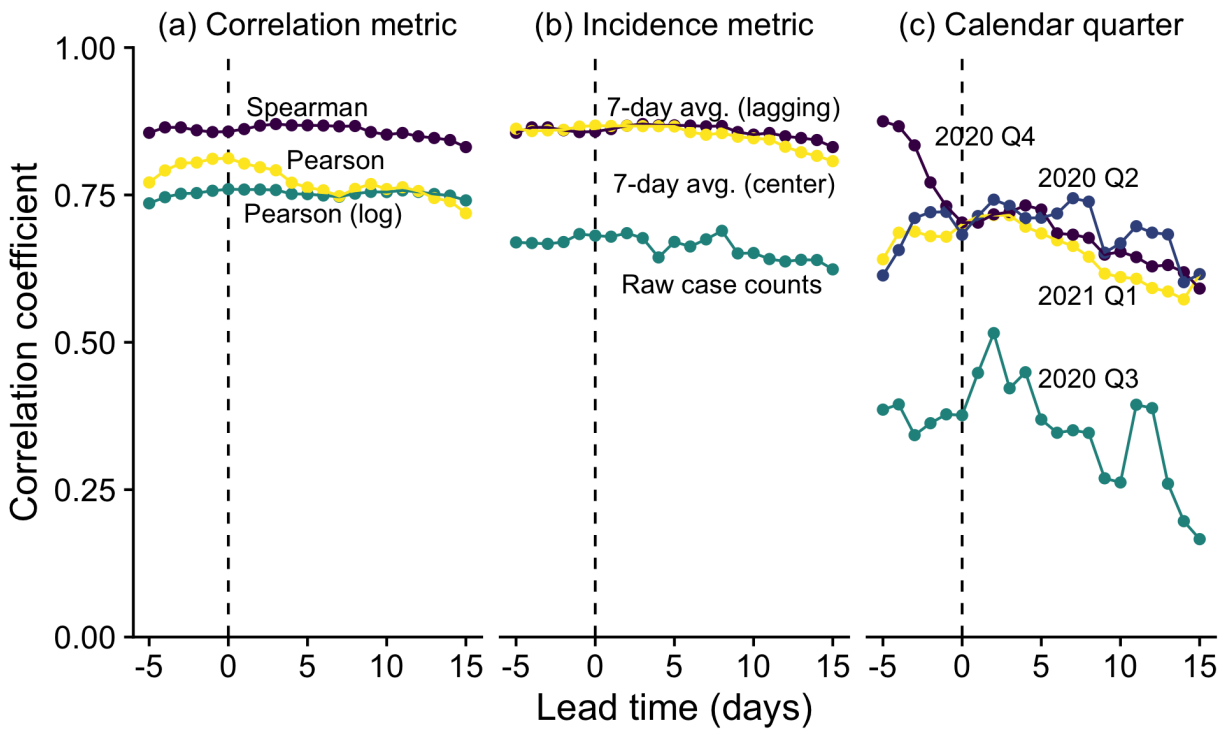

(a) Spearman: Spearman correlation between 7-day average case counts (i.e., mean number of new case counts over the day of wastewater sampling and the previous 6 days) and wastewater virus concentrations. This is the baseline analysis.

(a) Pearson: Pearson correlation between 7-day average case counts and wastewater virus concentrations.

(a) Pearson (log): Pearson correlation between 7-day average case counts and base-10 logarithms of wastewater virus concentrations.

(b) 7-day average (lagging): Baseline analysis; same as (a) Spearman.

(b) 7-day average (center): Spearman correlation between centered 7-day average case counts (i.e., mean number of new case counts over the day of wastewater sampling, the preceding 3 days, and the following 3 days) and wastewater virus concentrations.

(b) Raw case counts: Spearman correlation between daily number of new case counts (without smoothing) and wastewater virus concentrations.

(c) Baseline analysis like in (a) Spearman, but using data only from each calendar quarter (e.g., 2020 Q2 is April-June 2020).